\documentclass[preprint,aps,nofootinbib]{revtex4}

\newcommand {\bc}{\begin{center}}
\newcommand {\ec}{\end{center}}
\newcommand {\bea}{\begin{eqnarray}}
\newcommand {\eea}{\end{eqnarray}}
\newcommand {\be}{\begin{equation}}
\newcommand {\ee}{\end{equation}}

\def\lsim{\mathrel{\rlap{\lower4pt\hbox{$\sim$}}
    \raise1pt\hbox{$<$}}}               
\def\gsim{\mathrel{\rlap{\lower4pt\hbox{$\sim$}}
    \raise1pt\hbox{$>$}}}  
\usepackage{graphicx}

\begin{document}


\title{Critical behavior of the bulk viscosity in QCD}

\author{M.~Martinez$^1$, T.~Sch\"afer$^1$, and V.~Skokov$^{1,2}$}

\affiliation{$^1$Department of Physics, North Carolina State University,
  Raleigh, NC 27695\\
  $^2$RIKEN-BNL Research Center, Brookhaven National Laboratory, Upton,
  NY 11973}

\begin{abstract}
  We study the behavior of the bulk viscosity $\zeta$ in QCD near a 
possible critical endpoint. We verify the expectation that $(\zeta/s)\sim 
a(\xi/\xi_0)^{x_\zeta}$, where $s$ is the entropy density, $\xi$ is the 
correlation length, $\xi_0$ is the non-critical correlation length, 
$a$ is a constant and $x_\zeta\simeq 3$. Using a recently developed 
equation of state that includes a critical point in the universality 
class of the Ising model we estimate the constant of proportionality 
$a$. We find that $a$ is typically quite small, $a\sim O(10^{-4})$. We 
observe however, that this result is sensitive to the commonly
made  assumption that the Ising temperature axis is approximately 
aligned with the QCD chemical potential axis. If this is not the 
case, then the critical $\zeta/s$ can approach the non-critical 
value of $\eta/s$, where $\eta$ is the shear viscosity, even if the 
enhancement of the correlation length is modest, $\xi/\xi_0\sim 2$. 
\end{abstract}

\maketitle

\section{Introduction}
\label{sec_intro}

There are several programs dedicated to exploring the phase diagram
of Quantumchromodynamics (QCD) at heavy ion accelerator laboratories
around the world \cite{Braun-Munzinger:2015hba}. A central feature of
the phase diagram is a possible critical endpoint of a first order phase
transition between the hadronic phase and the quark gluon plasma (QGP)
phase. Experimentally, a critical point is expected to manifest itself
in terms of a non-trivial beam energy, rapidity, or system size
dependence of fluctuation observables \cite{Stephanov:1998dy}.

In a static system fluctuation observables are controlled by the critical
equation of state \cite{Stephanov:2008qz,Asakawa:2009aj,Stephanov:2011pb}.
The equation of state near the QCD critical point is expected to be
in the liquid-gas (Ising) universality class. The equation of state of
the Ising model is known from lattice simulations \cite{Engels:2002fi},
and accurate parametrizations are available in the literature
\cite{Zinn-Justin:2002ru}. More recently, there have been efforts to
map the Ising equation of state onto the QCD phase diagram, taking into
account information from lattice QCD about the equation of state and
the susceptibilities at zero baryon chemical potential
\cite{Nonaka:2004pg,Parotto:2018pwx}.

The dynamic behavior of fluctuations is expected to be governed by
model H in the classification of dynamical critical phenomena by
Hohenberg and Halperin \cite{Son:2004iv,Hohenberg:1977ym,Folk:2006ve}. 
Model H is a hydrodynamic theory that describes to coupling of the order
parameter field to a conserved momentum density. It predicts the
dynamic critical exponent for the relaxation of the order parameter,
and the critical behavior of the transport coefficients, the shear
viscosity $\eta$, the bulk viscosity $\zeta$, and the thermal
conductivity $\kappa$. In model H fluctuations of the order
parameter with wave number $q\sim\xi^{-1}$, where $\xi$ is the
correlation length, relax on a time scale $\tau\sim \xi^z$, where
$z\simeq 3$. This behavior is intermediate between ordinary
diffusion ($z\simeq 2$), and critical relaxation of a conserved charge not
coupled to fluctuations of the fluid velocity ($z\simeq 4$).
There is a very mild divergence in the shear viscosity, and more
pronounced critical behavior in the thermal conductivity and
bulk viscosity \cite{Kadanoff:1968,Hohenberg:1977ym,Kroll:1981,Ferrell:1985,Onuki:1997,Onuki:2002},
\be
\label{dyn_crit}
\eta\sim \xi^{0.05}\, , \hspace{1cm}
\kappa\sim \xi^{0.9}\, , \hspace{1cm}
\zeta\sim \xi^{2.8}\, . 
\ee
Physical effects related to critical transport phenomena have
been observed in ordinary fluids. For example, the critical behavior 
of the bulk viscosity manifests itself in sound attenuation near
the critical endpoint \cite{Onuki:2002}.

Recently, a number of authors have investigated the dynamic
evolution of fluctuations in an expanding QCD medium. This
includes studies of non-critical correlation functions
\cite{Akamatsu:2016llw,Akamatsu:2017rdu,Martinez:2018wia}, simulations of 
critical stochastic diffusion in an expanding medium \cite{Nahrgang:2018afz},
and deterministic frameworks for the evolution of two-point
\cite{Stephanov:2017ghc,Akamatsu:2018vjr,An:2019osr} or higher
n-point functions \cite{Mukherjee:2015swa} near a critical point.

There is a general expectation that the large critical exponent in
equ.~(\ref{dyn_crit}), combined with the strong deviation of the QCD
equation of state from scale invariance, will lead to a substantial
enhancement of the bulk viscosity, and to large effects on the
evolution of a heavy ion collision near a critical end point
\cite{Karsch:2007jc,Monnai:2016kud}. Our goal in the present work is
to study this problem more quantitatively, based on the equation of
state constructed by Parotto et al.~\cite{Parotto:2018pwx}. we will
verify the expected scaling behavior, and estimate the overall
coefficient. We will also study the relaxation of the bulk pressure
near a QCD critical point. These results complement earlier studies
of non-critical contributions to the bulk viscosity in QCD
\cite{Arnold:2006fz,Moore:2008ws,Lu:2011df,Dusling:2011fd}.

\section{Fluctuations of the energy density and pressure}
\label{sec_delta_p}

\begin{figure}[t]
\bc\includegraphics[width=8cm]{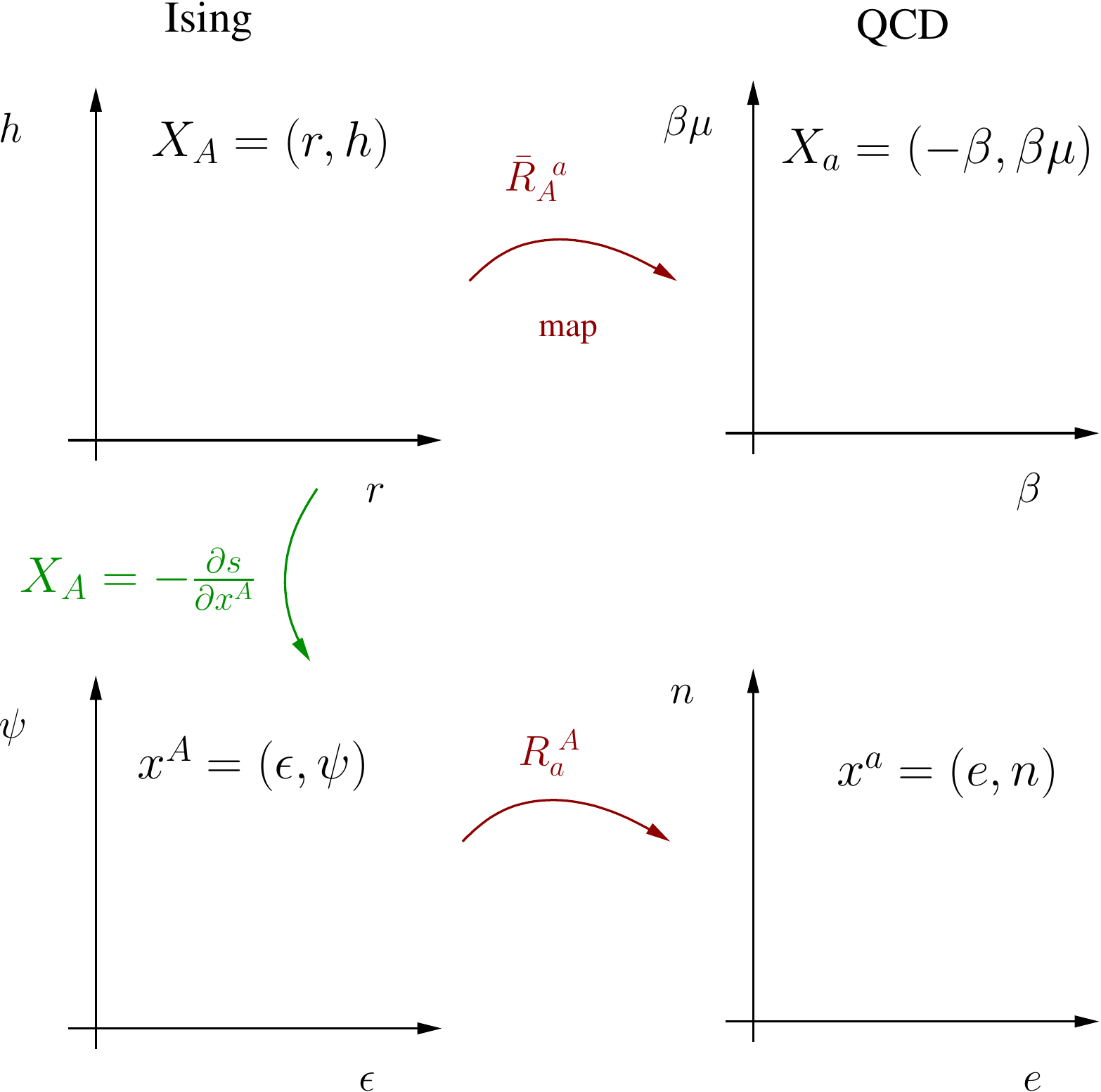}\ec
\caption{\label{fig_map}
Map from the Ising model to QCD for intensive variables and conserved
densities. Thermodynamic variables are defined in the text. The maps
$\bar{R}^b_A$ and ${R}^A_b$ are defined in equ.~(\ref{R_b_A},\ref{R_A_b}).
}   
\end{figure}

Fluctuations of the order parameter are governed by an entropy
functional $S=\int d^3x\, s$. The entropy functional of the Ising 
model is a function of the densities
\be
x^A=(\epsilon,\psi)\, ,
\ee
where $\epsilon$ and $\psi$ are the Ising energy density and order 
parameter. The corresponding intensive variables are the reduced 
temperature $r$ and the magnetic field $h$, 
\be
X_A= - \frac{\partial s}{\partial x^A} = (r,h)\, .
\ee
We are following the notation of Landau and Lifshitz \cite{Landau:smI} 
as well as Akamatsu et al.~\cite{Akamatsu:2018vjr}. The analogous 
canonical pair in QCD is
\be
x^a=(e,n)\, ,  \hspace{1cm}
X_a=(-\beta,\beta\mu)\, , 
\ee
where $(e,n)$ are the energy and baryon density, $\beta=1/T$ is the 
inverse temperature and $\mu$ is baryon chemical potential. We will 
assume that there is a map between the intensive variables in QCD and 
the corresponding Ising variables, see Fig.~\ref{fig_map} and 
Sect.~\ref{sec_entropy}. Fluctuations of the QCD pressure are given by 
\be
\label{del_P}
 \delta P = \left.\frac{\partial P}{\partial\beta}\right|_{\beta\mu}
            \delta\beta
   + \left.\frac{\partial P}{\partial\beta\mu}\right|_{\beta}
   \delta(\beta\mu)
  = -\frac{e+P}{\beta}\,\delta\beta + \frac{n}{\beta}\, \delta(\beta\mu) \, . 
\ee
Using the fact that $(\beta,\beta\mu)$ is conjugate to $(e,n)$
we obtain
\be
\label{del_P_2}
 \delta P = \frac{e+P}{\beta}\frac{\partial s}{\partial(\delta e)} 
           - \frac{n}{\beta} \frac{\partial s}{\partial(\delta n)}.
\ee
The map between $(r,h)$ and $(\beta,\beta\mu)$ induces a map between
the QCD densities $(e,n)$ on the Ising densities $(\epsilon,\psi)$.
We assume that the singular part of the QCD entropy density is proportional
to the Ising entropy density, $s^{\it sing}(e,n)=As^{\it Is}(\epsilon(e,n),
\psi(e,n))$. A common assumption is that the images of the $r$ and $h$
axes in the QCD phase diagram are approximately orthogonal, and that the
Ising temperature axis is almost aligned with the QCD chemical potential
axis\footnote{We will reconsider this assumption in Sect.~\ref{sec_kaw}.}, 
$\partial\epsilon/\partial e\simeq 0$
\cite{Nonaka:2004pg,Bluhm:2016byc,Parotto:2018pwx}. This implies that  
\be
\label{del_P_3}
 \delta P = \frac{e+P}{\beta}\frac{\partial \psi}{\partial (\delta e)}
\frac{\partial s^{\it Is}}{\partial\psi} 
           - \frac{n}{\beta} \frac{\partial \epsilon}{\partial (\delta n)}
     \frac{\partial s^{\it Is}}{\partial\epsilon}.
\ee
Onuki observed that the Ising entropy functional contains a tri-linear
coupling between $\epsilon$ and $\psi^2$, and as a result fluctuations
of the pressure couple to $\psi^2$ \cite{Onuki:1997,Onuki:2002}. This
means that correlation functions of $\delta P$ are controlled by the
order parameter relaxation rate, and the slow relaxation of $\psi$
leads to a large enhancement in the bulk viscosity. 

The presence of a tri-linear coupling between $\epsilon$ and
$\psi^2$ in the Ising entropy $S[\psi,\epsilon]$ can be understood
from the relation between entropy and Gibbs free energy\footnote{Note
that there is some disagreement in the literature on whether $G[\psi,r]$ 
should be called free energy or Gibbs free energy. Zinn-Justin refers to this
functional as a free energy and equ.~(3) of Asakawa et al.~\cite{Nonaka:2004pg}
as well as Parotto et al.~\cite{Parotto:2018pwx} follows that notation,
but Akamatsu et al.~\cite{Akamatsu:2018vjr} uses the term Gibbs free
energy, and reserves the term free energy for $F[h,r]$.} 
\cite{Halperin:1974} 
\be
\exp(-\beta G[\psi,r]) =
\int {\cal D\epsilon} \exp(S[\psi,\epsilon]-E/T)\, ,
\hspace{0.5cm}
E=\int d^3x\, (\epsilon+\epsilon_0)\, ,
\ee
where $E$ is the total energy, $\epsilon_0$ is the background energy
density, and $T$ is the temperature. For small $\epsilon$ and $\psi$
we can expand the entropy functional as
\be
\label{S_ising}
S[\psi,\epsilon]=-\int d^3x\, \left\{
k (\nabla\psi)^2 + \frac{v}{2}\psi^2 + \frac{u}{4}\psi^4
+\gamma\epsilon\psi^2 + \frac{1}{2C_0}\epsilon^2 \right\}
+ S_0+\frac{E}{T_0}\, , 
\ee
where $k,v,u$ and $\gamma$ are constants. $C_0$ is the
heat capacity, and $S_0$ and $T_0$ are the background entropy and
temperature. The free energy functional is
\be
\beta G[\psi,r]=\int d^3x\, \left\{
k (\nabla\psi)^2 + \frac{\tilde{v}}{2}\psi^2
+ \frac{\tilde{u}}{4}\psi^4 \right\}
\ee
and $\tilde{v}=v+2\gamma C_0(T-T_0)/T_0^2$, so the Legendre transform
generates the $T$ dependence of the correlation length. Note that in
this Gaussian approximation $\gamma$ is a constant, independent of $T$.
We will see in the following Section that, in general, $\gamma$
scales with $T-T_0$ according a non-trivial critical exponent.

Finally, combining equ.~(\ref{del_P_3}) and equ.~(\ref{S_ising})
we obtain the slow mode contribution to fluctuations of the
pressure
\be
\label{del_P_psi}
 \delta P = nT A a_{n\epsilon}\gamma\psi^2
\ee
where we have defined $a_{n\epsilon}=(\partial\epsilon)/(\partial (\delta n))$.

\section{Critical entropy functional}
\label{sec_entropy}

We determine the critical entropy density using the Ising equation of
state constructed by Zinn-Justin \cite{Zinn-Justin:2002ru}. We write
the Gibbs free energy as \cite{Schofield:1969,Zinn-Justin:2002ru}
\be
G[\psi,r] = h_0M_0 R^{2-\alpha} g(\theta)\, , 
\ee
where $r=(T-T_c)/T_c$. The constants $h_0$ and $M_0$ will be specified
below, $\alpha= 0.11$ is the specific heat exponent, and $g(\theta)$ is a
function that is given in App.~\ref{sec_ising_eos}. We use the $(R,\theta)$
coordinates
\bea
\psi &=& M_0R^\beta \theta \, ,\\
r    &=&  R(1-\theta^2)\, ,
\eea
where $\beta=0.33$ is the order parameter exponent. Here, $R\in[0,\infty)$
and $\theta\in[-\theta_0,\theta_0]$, where $\theta_0$ is determined
by the condition $\tilde{h}(\theta_0)=0$. Furthermore, $\tilde{h}(\theta)$ 
is a function that appears in the magnetic equation of state 
\be
h=h_0R^{\beta\delta}\tilde{h}(\theta)\, , 
\ee
$\delta=4.78$ is the external field exponent, and $\tilde{h}(\theta)$
is also specified in App.~\ref{sec_ising_eos}. For this parameterization we
find $\theta_0=1.154$. The value of $\theta_0$ determines the boundaries of
the first order region for $r<0$, $\psi_\pm=\pm M_0R^\beta\theta_0$. 

We can expand the Gibbs free energy for small $r$ and $\psi$. We
find
\be
G[\psi,r]\simeq h_0M_0\left\{
g_\pm r^{2-\alpha}+\frac{1}{2}m^2_\pm r^{2-\alpha-2\beta}
(\psi-\psi_0)^2 + \ldots \right\} \, ,
\ee
where 
\be
g_\pm = \left\{ \begin{array}{cc}
  -0.84   & r>0 \\
  -1.58   & r<0
\end{array}\right.
\hspace{1cm}
m^2_\pm = \left\{ \begin{array}{cc}
  1.00   & r>0 \\
  4.77   & r<0
\end{array}\right.
\hspace{1cm}
\psi_0 = \left\{ \begin{array}{cc}
    0        & r>0 \\
  \psi_\pm   & r<0
\end{array}\right.
\ee
Note that in the mean field approximation we have $\alpha=0$ and
$\beta=1/2$, so that $G\sim \frac{1}{2}r\psi^2$, as expected. Also note 
that $2-\alpha-2\beta=\tilde{\gamma}$ is the susceptibility exponent.
The Ising energy density is
\be
\epsilon = \frac{\partial G}{\partial r}
  \simeq h_0M_0\left\{
g_\pm (2-\alpha)r^{1-\alpha}+\frac{1}{2}m^2_\pm (2-\alpha-2\beta)
  r^{1-\alpha-2\beta} (\psi-\psi_0)^2 + \ldots \right\} \, ,
\ee
We can now determine an entropy functional that describes 
fluctuations of the energy density at constant $\psi$ by  
performing a Legendre transformation, $s[\psi,\epsilon]=
G[\psi,r]-r\epsilon$. Expanding $s$ for small values of 
the arguments gives
\bea
s[\psi,\epsilon] &\simeq& (-g_\pm)(1-\alpha)
  \, {\cal E}^{\frac{2-\alpha}{1-\alpha}}
  + \frac{1}{2}m_\pm^2
  \, {\cal E}^{\frac{2-\alpha-2\beta}{1-\alpha}}(\psi-\psi_0)^2
   + \ldots\, , \\
   &&    \hspace{0.5cm}
{\cal E} = \frac{\epsilon}{(-g_\pm)(2-\alpha)\, h_0M_0}\,  .
   \nonumber
\eea
We can now read off the tri-linear coupling defined in the previous
section. We get
\be
\label{gamma_pm}
\gamma_\pm =  \frac{m_\pm^2}{2(-g_\pm)}
    \frac{2-\alpha-2\beta}{(2-\alpha)(1-\alpha)}\,
    |r|^{1-2\beta}\, .
\ee
There are two differences compared to the mean field result in the 
previous section. The first is that the tri-linear coupling vanishes 
near the transition point, with a critical scaling controlled by the 
exponent\footnote{Note that this exponent agrees with the exponent determined 
using renormalization group arguments \cite{Halperin:1974,Onuki:1997}. 
Halperin et al.~provide a diagrammatic argument that explains why the 
critical scaling of $\gamma$ is related to the specific heat and 
susceptibility exponents.}
$(1-2\beta)\simeq 0.34$. We will see that this is more than 
compensated by the divergence in the order parameter relaxation rate. 
The second is that there is an amplitude ratio, that means the coupling 
is different on the first order ($r<0$) and crossover ($r>0$) of the 
transition. We find
\be
\label{gamma_pm_2}
\gamma_\pm = \left\{ \begin{array}{cc}
    0.43\,  r^{1-2\beta}     & r>0 \\
    1.10\, |r|^{1-2\beta}    & r<0
\end{array}\right.
\ee
Note that the scaling with $r$ can be converted to a scaling 
relation involving the correlation length using $|r|\sim \xi^{-1/\nu}$,
with $\nu\simeq 0.63$. Indeed, Zinn-Justin provides a scaling form
of the correlation length
\be 
 \xi = \xi_0R^{-\nu}g_\xi^{1/2}(\theta)\, , 
\ee 
where $\xi_0$ is an overall scale and $g_\xi(\theta)\simeq (1-5\theta^2
/18)$ \cite{Zinn-Justin:2002ru}.

Parotto et al.~construct a map $X_A(X_b)$ from intensive
QCD variables $X_a=(-\beta,\beta\mu)$ to Ising variables $X_A=(r,h)$.
The specific map considered in \cite{Parotto:2018pwx} is a simple
linear relation
\bea
\label{map_par_1}
\frac{T-T_c}{T_c} &=& \bar{w}\left(
r\bar{\rho}\sin(\alpha_1) + h\sin(\alpha_2)\right) \, , \\
\label{map_par_2}
\frac{\mu-\mu_c}{T_c} &=& \bar{w}\left(
-r\bar{\rho}\cos(\alpha_1) - h\cos(\alpha_2)\right) \, , 
\eea
where $\bar{w},\bar{\rho}$ and $\alpha_{1,2}$ are parameters.
Most of the work in the existing literature
\cite{Nonaka:2004pg,Parotto:2018pwx,Bluhm:2016byc} assumes
that $\alpha_1\simeq 0$ and $\alpha_2\simeq \pi/2$. We will
discuss this assumption in more detail below, but for now
we will assume that $\alpha_1= 0$ and $\alpha_2= \pi/2$. Parotto
et al.~choose $\bar{w}=1,\bar{\rho}=2$ and $A=T_c^3$. They 
also use $M_0=0.605$ and $h_0=0.394$.

In order to determine the map between the conserved densities we
use the fact that the transformation between the intensive variables
can be specified in terms of the matrix (see Fig.~\ref{fig_map})
\be
\label{R_b_A}
\bar{R}^b_A=\frac{\partial X_A}{\partial X_b} .
\ee
This map induces a relationship between the densities $x_A(x_b)$,
described by
\be
\label{R_A_b}
{R}^A_b=\frac{\partial x^A}{\partial x^b} .
\ee
Since the densities are conjugate to the intensive variables we
must have
\be
\label{R_Rbar}
R^A_b\bar{R}^b_C= \delta^A_C\, ,
\ee
which implies that the matrix $R$ is the inverse of $\bar{R}$.
In the simple case that $\alpha_1=0$ and $\alpha_2=\pi/2$ we
obtain
\be
\label{a_ne}
a_{n\epsilon} = \frac{\partial\epsilon}{\partial (\delta n)}
  = \frac{\bar{\rho}\bar{w}}{A} \, 
\ee
where we have used that the normalization of the Ising and QCD
entropy functional differ by a factor $A$.

\section{Critical bulk viscosity}
\label{sec_bulk}

 The bulk viscosity is determined by the Kubo relation
\be
 \label{zeta_Kubo}
 \zeta(\omega) = -\frac{1}{9\omega}{\rm Im}\, G^{ii,jj}_R(\omega,0)
\ee
where $G_R^{ij,kl}(\omega,k)$ is the retarded correlation 
function of the (spatial components) of the stress tensor
\be
\label{G_ret}
G_R^{ij,kl}(\omega,k) = -i 
 \int dtd^3x\, e^{-i(\omega t-kx)}\langle[\Pi^{ij}(0,0),\Pi^{kl}(x,t)]\rangle
  \theta(t)\, . 
\ee
The critical behavior arises from the contribution of the non-equilibrium
pressures to the trace of the stress tensor, $\frac{1}{3}\Pi^{ii}=\delta P$,
with $\delta P$ given in equ.~(\ref{del_P_psi}). In order to compute
the right hand side of the Kubo relation equ.~(\ref{zeta_Kubo}) we
consider the symmetric correlation function
\be
\label{G_S}
G_S(\omega,k) = c
\int dtd^3x e^{-i(\omega t-kx)}\langle \psi^2(0)\psi^2(t,x) \rangle
\ee
where we have defined $c=(nT A a_{n\epsilon}\gamma)^2$. In statistical
field theory we can factorize the correlation function and determine
the retarded correlator using the fluctuation-dissipation relation.
We get
\be
 \label{G_R}
G_R(\omega,k=0) = 2c
\int \frac{d\omega'}{2\pi} \int \frac{d^3k}{(2\pi)^3}
\left\{\Delta_R(\omega-\omega',k)\Delta_S(\omega',k)
+ (S\leftrightarrow R)\right\}\, ,   
\ee
where $\Delta_{S,R}$ are the symmetric and retarded correlation
function of the order parameter $\psi$. The slow dynamics of $\psi$
is governed by a diffusion equation \cite{Hohenberg:1977ym}
\be
\label{mod_H}
\frac{\partial\psi}{\partial t}  =
\lambda_0\nabla^2 \frac{\delta G}{\delta \psi} +\theta.
\ee
Here, $\lambda_0$ is a transport coefficient and $\theta$ is a
stochastic force. The noise correlator is
\be
\label{noise}
\langle \theta(x,t)\theta(x',t')\rangle
= -2\lambda_0T \nabla^2\delta^3(x-x')\delta(t-t')\, . 
\ee
The diffusion constant is $D=\lambda_0/\chi_0$ where the
susceptibility is given by 
\be
\label{suscep}
\chi_k = \frac{\xi^2}{1+(k\xi)^2}. 
\ee
The retarded and symmetric order parameter correlation
functions are
\bea
\label{Del_R}
\Delta_R(\omega,k) &=&  \chi_k \, \frac{\Gamma_k}{-i\omega + \Gamma_k}\, ,  \\
\label{Del_S}
\Delta_S(\omega,k) &=&  2\chi_k T\, \frac{\Gamma_k}{\omega^2+\Gamma_k^2}
  \, , 
\eea
where $\Gamma_k=\lambda_0k^2/\chi_k$. The bulk viscosity is
\be
\label{zeta_w}
\zeta(\omega) = c \int \frac{d^3k}{(2\pi)^3} \frac{2T\chi_k^2}
     {-i\omega + 2\Gamma_k}\, . 
\ee
We note that the order parameter field $\psi$ is dimensionless. We
can absorb all dimensionful parameters in equ.~(\ref{mod_H}) into a
length scale, the non-critical correlation length $\xi_0$, and a
non-critical relaxation time $t_0$. These are two free parameters
in the model H description of the QCD critical point, in addition
to the parameters that appear in the equation of state. As a rough
estimate of $\xi_0$ we will use the entropy density of the system,
and assume that $s\xi_0^3=1$. We will comment on this choice in
Sect.~\ref{sec_out}. The relaxation time is determined by the 
diffusion constant, $D_0=t_0/\xi_0^2$.

The integral in equ.~(\ref{zeta_w}) is easily performed in
the $\omega\to 0$ limit. We find
\be 
\label{zeta_est_1}
\frac{\zeta}{s} = \left(\gamma^R_\pm a_{n\epsilon}\right)^2
\left(t_0T\right)^2 \left(\frac{n}{s}\right)^2
\frac{3}{32\pi} \left( \frac{\xi}{\xi_0} \right)^{3.92}\, , 
\ee
where we have taken into account the scaling of $\gamma_\pm\equiv
\gamma^R_\pm r^{1-2\beta}$, see equ.~(\ref{gamma_pm_2}). The critical
exponent $x_\zeta=3.92$ is larger than the one found by Onuki, see 
equ.~(\ref{dyn_crit}). We will show in Sect.~\ref{sec_kaw} that this
is due to the fact that in equ.~(\ref{mod_H}) we have neglected 
the coupling of the order parameter to the momentum density 
of the fluid. The coupling $\gamma^R_\pm a_{n\epsilon}$ was 
determined in equ.~(\ref{gamma_pm_2}) and (\ref{a_ne}). For typical
values of the diffusion constant $t_0 T$ is of order one; we have 
used $t_0\simeq 1.8$ fm from \cite{Akamatsu:2018vjr}. The coefficient
$3/(32\pi)\sim 3\cdot 10^{-2}$ is a loop factor. The main uncertainty
in equ.~(\ref{zeta_est_1}) is the factor $n/s$. For typical
equations of state that put the critical endpoint within reach
of the RHIC beam energy scan program this coefficient is quite 
small. For example, the equation of state shown in Fig.~6 of 
Parotto et al.~\cite{Parotto:2018pwx} has $n/s\simeq 1/25$. We
then get
\be
\label{zeta_est_2}
\frac{\zeta}{s} = \left(\frac{\xi}{\xi_0}\right)^{3.92}
  \left\{ \begin{array}{cc}
    6.6\cdot 10^{-5}\;  & r>0  \\
    4.2\cdot 10^{-4}\;  & r<0  
\end{array}\right. \, . 
\ee
We conclude that the critical enhancement in the bulk viscosity is 
not large, unless the correlation length becomes quite large. We 
will study the growth in the effective correlation length in the 
Sect.~\ref{sec_exp}, and we will consider a possible scenario that 
overcomes the suppression by $n/s$ in the following Section.

\begin{figure}[t]
\bc\includegraphics[width=8.5cm]{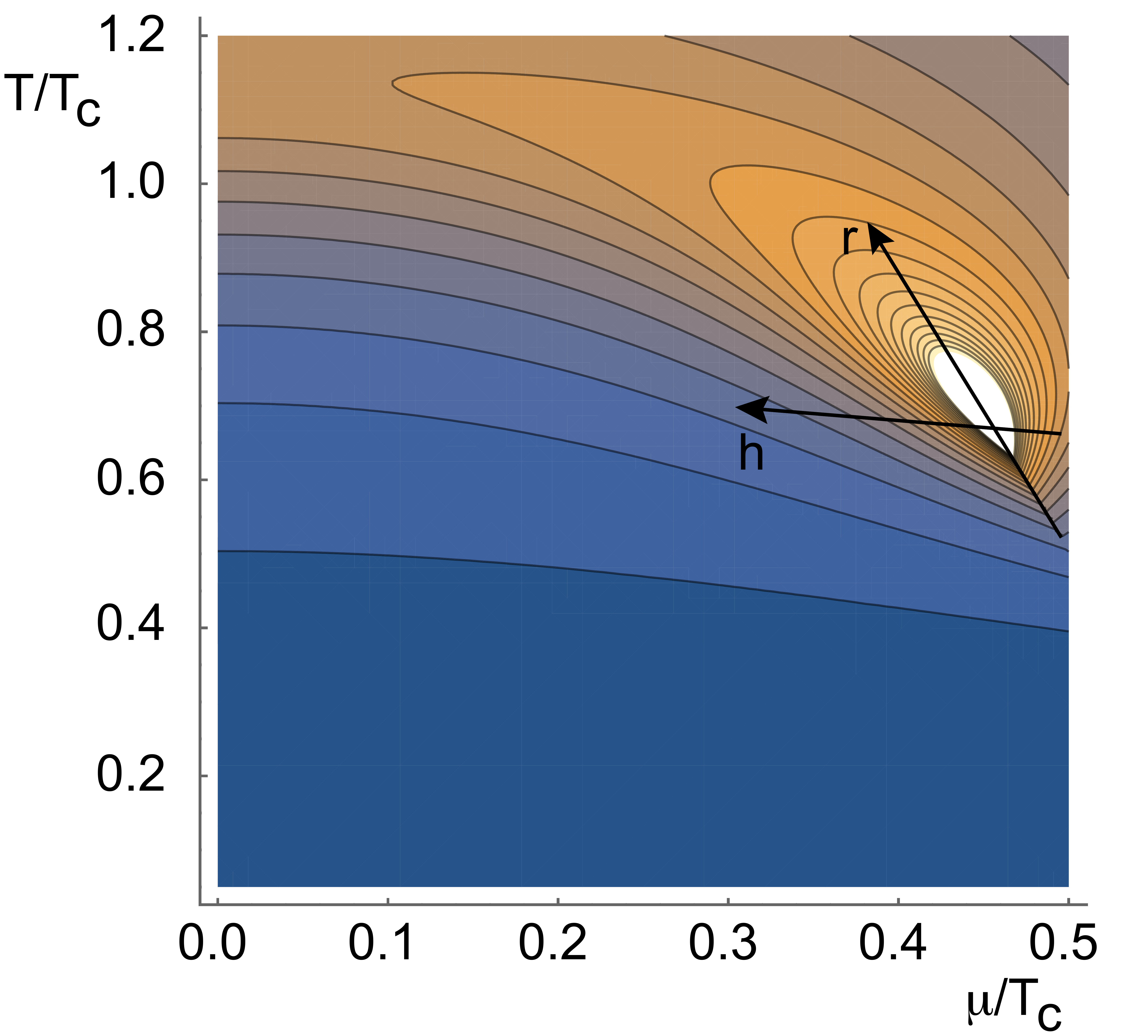}\ec
\caption{\label{fig_rmt}
  Ising model axes $r$ and $h$ in a random matrix model of the QCD
phase diagram. The axes of the plot are the quark chemical potential
$\mu$ and the temperature $T$ in units of the critical temperature
in the chiral limit. The parameters of the model were chosen as
explained in the text. The contour lines show the chiral
susceptibility. 
}   
\end{figure}

\section{Refinements}
\label{sec_kaw}

 In this section we will discuss two refinements to the 
calculation described in the previous sections. The first involves
the treatment of the order parameter relaxation rate. In model H
it has been shown that the relaxation rate $\Gamma_k$ is governed
by the coupling of $\psi$ to the momentum density of the fluid
\cite{Hohenberg:1977ym}. Consider, for example, the choice $\psi=s/n$.
Then the conservation laws generate a coupling of the form $(\partial 
\psi)/(\partial t)=\vec\pi\cdot\vec{\nabla}\psi+\ldots$, where $\vec{\pi}$
is the momentum density of the fluid. The momentum relaxation rate has a 
very weak divergence, so that it makes sense to approximate that rate
by the non-critical shear viscosity $\eta_0$ of the fluid. This is known 
as the Kawasaki approximation. Within this approximation
\cite{Kawasaki:1970,Hohenberg:1977ym,Onuki:2002}
\be
\label{Gam_Kaw}
\Gamma_k = \frac{T\xi^{-3}}{6\pi\eta_0} \, K(k\xi)
\ee
where 
\be
\label{Kaw_fct}
K(x) = \frac{3}{4}\left[ 1 + x^2 + \left(x^3+x^{-1}\right) 
 \arctan(x)\right]\, . 
\ee
We will also use a refined parameterization of the order 
parameter susceptibility 
\be 
\label{chi_crit}
  \chi_k=\frac{\chi_0}{1+(k\xi)^{2-\eta}}
\ee
where $\eta\simeq 0.036$ is the correlation function exponent and 
$\chi_0=\xi_0^2 (\xi/\xi_0)^{2-\eta}$. In this context we will continue
to define $\xi_0$ by $(s\xi_0^3)=1$, but we will take $t_0=c\,\xi_0$, 
where $c$ is the speed of light, and the relaxation time scale is set 
by the non-critical value of $\eta_0/s$. We can now recompute the 
critical contribution to the bulk viscosity in equ.~(\ref{zeta_w}). 
We find 
\be 
\label{zeta_est_3}
\frac{\zeta}{s} = \left(\gamma^R_\pm a_{n\epsilon}\right)^2
\left(t_0T\right) \left(\frac{n}{s}\right)^2
\frac{3}{4\pi^2}\left( 4\pi\frac{\eta_0}{s}\right) I_K
 \left( \frac{\xi}{\xi_0} \right)^{z-\alpha/\nu}\, ,
\ee
where 
\be
I_K=\int dy\, \frac{y^2}{K(y)(1+y^{2-\eta})^2}\simeq 0.649
\ee 
is the result of a numerical integral, and
we have used the hyperscaling relation $\nu d=2-\alpha=2\beta+\gamma$, 
where $d=3$ is the number of spatial dimensions. For the Kawasaki 
function we have $z=3$, which is close to the result in the 
$\epsilon$-expansion, $z=3.05$. Using the values of $\alpha$ and $\nu$ 
quoted above we find $z-\alpha/\nu\simeq 2.8$, which agrees with 
Onuki's result \cite{Onuki:1997}. For $\eta_0/s\simeq 1/(4\pi)$ the 
estimate in equ.~(\ref{zeta_est_3}) is numerically close to the 
result given above. We find
\be
\label{zeta_est_4}
\frac{\zeta}{s} = \left(4\pi\frac{\eta_0}{s}\right)
   \left(\frac{\xi}{\xi_0}\right)^{2.8}
    \left\{ \begin{array}{cc}
    8.0\cdot 10^{-5}\;   & r>0  \\
    5.2\cdot 10^{-4}\;   & r<0  
\end{array}\right. \, . 
\ee
We note that the smallness of the pre-exponent in equ.~(\ref{zeta_est_4})
is mainly due to the small factor $(n/s)$. The appearance of this factor
is due to the assumption about the orientation of the Ising axes in the 
QCD phase diagram made above equ.~(\ref{del_P_3}). It was recently observed
that close to the chiral limit this assumption is not correct, and that 
in models of QCD this observation also applies to realistic values of 
the quark masses \cite{Pradeep:2019ccv}. To illustrate this fact we show in 
Fig.~\ref{fig_rmt} the critical endpoint and the orientation of the Ising 
axes in the random matrix model introduced in \cite{Halasz:1998qr}. The 
overall scale was fixed by the requirement that $T_\chi$, the phase transition 
temperature in the chiral limit, and $T_{pc}$, the the pseudo-critical
temperature for the physical value of the pion and kaon mass, is the 
same as in lattice QCD, $T_{\chi}/T_{pc}\simeq (132^{+3}_{-6}\,{\it MeV})/
(156.5 \pm 1.5\,{\it MeV})$ \cite{Ding:2019prx}. In the random matrix
model this correspond to a quark mass $m_q\simeq 10$ MeV. We find $\alpha_1
\simeq 105^\circ$ and $\alpha_2\simeq 165^\circ$. In this case the dominant
contribution to $\delta P$ arises from
\be 
\delta P = sT^2A a_{e\epsilon} \gamma\psi^2
\ee
with $a_{e\epsilon}=(\partial \epsilon)/(\partial (\delta e))$. Using 
the mapping relation in  equ.~(\ref{map_par_1},\ref{map_par_2})
together with equ.~(\ref{R_Rbar}) we find  $Aa_{e\epsilon}=\beta\bar{\rho}
\bar{w}\sin(\alpha_1)$ and 
\be
\label{zeta_est_5}
\frac{\zeta}{s} = \sin(\alpha_1)^2
 \left(4\pi\frac{\eta_0}{s}\right)
  \left(\frac{\xi}{\xi_0}\right)^{2.8}
\left\{ \begin{array}{cc}
    5.1\cdot 10^{-2} \;  & r>0  \\
    3.2\cdot 10^{-1} \;  & r<0  
\end{array}\right. \, .
\ee
With $\sin^2(\alpha_1)\simeq 1/4$ from Fig.~\ref{fig_rmt} we observe
that, at least on the first order side of the transition, the
critical bulk viscosity can reach $\zeta/s\sim 1/(4\pi)$, comparable
to the non-critical shear viscosity\footnote{Note that Parotto et
al.~\cite{Parotto:2018pwx} use $\alpha_1=3.85^o$, based on the 
curvature of the freezeout curve. This corresponds to $\sin(\alpha_1)^2
\simeq 4\cdot 10^{-3}$.}.
Note that equ.~(\ref{zeta_est_4},\ref{zeta_est_5}) are limiting cases,
corresponding to $\sin(\alpha_1)^2\sim 0$ and $\sin(\alpha_1)^2
\gsim (n/s)^2$, of a more general relation that involves both
$a_{n\epsilon}$ and $a_{e\epsilon}$. The more general result follows
from the coupling $\delta P=\bar{\rho}\bar{w}[s\sin(\alpha_1)
+ n(\cos(\alpha_1)-\beta\mu\sin(\alpha_1))]TA\gamma\psi^2$.

\section{Bulk viscosity in an expanding system}
\label{sec_exp}

 In a heavy ion collision the growth of the correlation length is limited 
by the failure of the evolution to go precisely through the critical 
point, and by the finite time available for the correlation length to 
reach its equilibrium value \cite{Berdnikov:1999ph,Akamatsu:2018vjr}. 
The recent study \cite{Akamatsu:2018vjr} concludes that the second effect 
dominates, and that $\xi/\xi_0$ is limited by $\xi_{KZ}/\xi_0$ where $\xi_{KZ}$ 
is the Kibble-Zurek scale. Akamatsu et al.~\cite{Akamatsu:2018vjr} 
estimate that  $\xi_{KZ}/\xi_0 \sim 1.33$, so that the critical bulk 
viscosity in equ.~(\ref{zeta_est_2}) does not become large. 

\begin{figure}[t]
\bc\includegraphics[width=8cm]{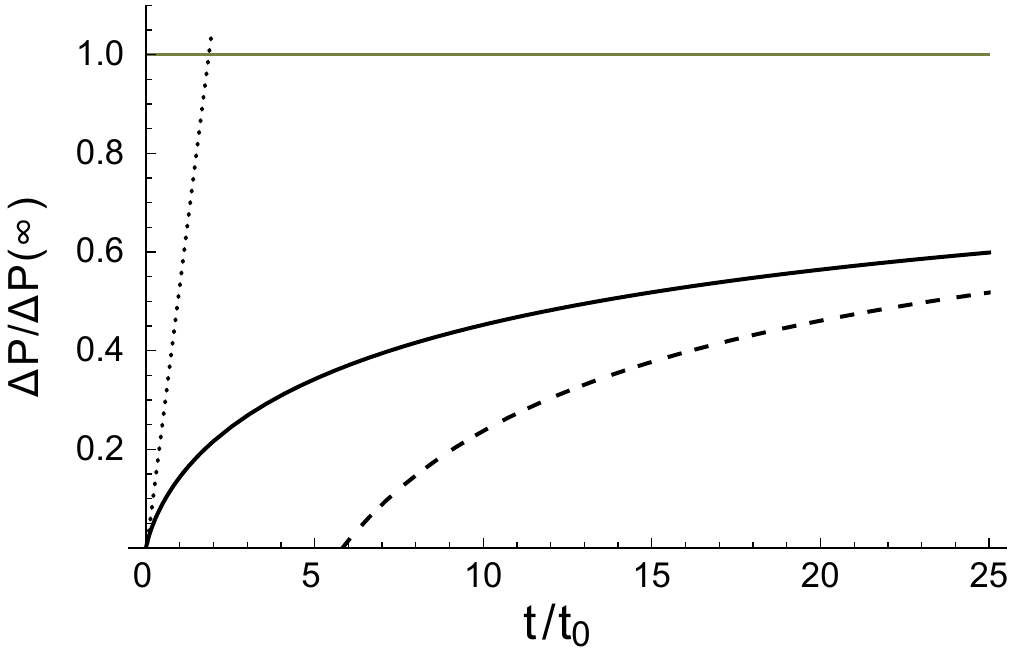}\ec
\caption{\label{fig_relax}
Relaxation of the bulk pressure $\Delta P(t)$ to its asymptotic
value $\Delta P(\infty)$ as a function of $t/t_0$, where $t_0$ is
the non-critical relaxation time. In this figure we have chosen
$\xi/\xi_0=1.5$. The dotted and dashed lines show the early and
late time asymptotics of $\Delta P(t)$.}   
\end{figure}

 There is a second effect which appears even if the correlation length
does become large. Critical slowing down also implies that the 
bulk pressure only relaxes slowly to the equilibrium value $\Delta P
\simeq \zeta(\vec{\nabla}\cdot \vec{u})$, where $\vec{u}$ is the fluid 
velocity \cite{Stephanov:2017ghc}. The response $\Delta P(t)$ to a 
time-dependent bulk stress $(\vec{\nabla}\cdot \vec{u})(t')$ is given 
by 
\be
\Delta P(t) = \int^t dt' \, G(t-t')(\vec{\nabla}\cdot \vec{u})(t')\, ,
\ee
where $G(t)$ is the Fourier transform of equ.~(\ref{zeta_w})
\be 
G(t) = c \int \frac{d^3k}{(2\pi)^3} \, 2T\chi_k^2 
     \exp\left(-2\Gamma_k t\right)\, . 
\ee
In the following, we will use the simple form $\Gamma_k=\lambda_0k^2
\chi_k^{-1}$. The result is easily generalized to the Kawasaki
approximation in equ.~(\ref{Gam_Kaw}).
The time integral of $G(t)$ is given by the bulk viscosity $\zeta$, and 
the asymptotic behavior of $G(t)$ for large $t$ is governed by hydrodynamic 
tails $G(t)\sim 1/t^{3/2}$ \cite{Alder:1967,Ernst:1971,Pomeau:1975,Kovtun:2003vj,Kovtun:2012rj,Martinez:2017jjf,Akamatsu:2016llw,Martinez:2018wia}.
The tail of the critical contribution is
\be 
\label{zeta_tail}
\frac{G(t)}{sT} = \left(\gamma^R_\pm a_{n\epsilon}\right)^2
\left(t_0T\right) \left(\frac{n}{s}\right)^2
\frac{1}{16\pi^{3/2}} \left( \frac{\xi}{\xi_0} \right)^{5.92}
   \left(\frac{t_0}{t}\right)^{3/2}\, , 
\ee
which has an even stronger dependence on the correlation length
than the static value of the bulk viscosity. We note, however, that 
the asymptotic behavior only sets in at parametrically late time,
$(t/t_0)\gg (\xi/\xi_0)^4$. 

 We may also consider a bulk stress that is turned on at $t=0$ and then
study the evolution of $\Delta P(t)$ towards its asymptotic value $\Delta P
(\infty)=\zeta (\vec{\nabla}\cdot \vec{u})$. The relaxation is not 
exponential. Instead, we find very slow relaxation 
\be 
\Delta P(t)\simeq \Delta P(\infty) 
\left\{ 1 - \frac{4}{3}\sqrt{\frac{2}{\pi}}
  \left(\frac{\xi}{\xi_0}\right)^2\left(\frac{t_0}{t}\right)^{1/2}
\right\}\, , 
\ee 
where, again, the asymptotic behavior requires $t\gg\xi^4$. 
For small $t$ we find that $\Delta P$ grows linearly in $t$, 
\be 
\Delta P(t) = 
 \frac{8}{3}\left(\frac{\xi_0}{\xi}\right)^4
 \left(\frac{t}{t_0}\right)
 \Delta P(\infty)\, ,
\ee  
but the linear growth involves a much smaller power of the
correlation length. This behavior is shown in Fig.~\ref{fig_relax}. 
We observe that even for a modest enhancement of the correlation length, 
$\xi/\xi_0=1.5$, the critical bulk pressure relaxes on a time scale 
much slower than the natural equilibration time $t_0$.

\section{Conclusions and outlook}
\label{sec_out}

In this work we have studied the critical bulk viscosity in QCD.
We find that the result is very sensitive to the orientation of the
Ising axes in the QCD phase diagram. It is often assumed that the
Ising temperature axis is aligned with the QCD chemical potential
axis, and that the Ising magnetic field axis is orthogonal to that. In
that case, the QCD bulk viscosity is suppressed by a factor $(n/s)^2$.
If we include a possible misalignment of the Ising axes our final
result is given in equ.~(\ref{zeta_est_5}). We observe that there 
is significant dependence on what side of the phase transition we
are considering; the critical contribution to the bulk viscosity is
about an order of magnitude larger on the first order side of the 
transition. In this regime the critical enhancement of the bulk 
viscosity may become comparable to the non-critical shear viscosity, 
even for a modest enhancement of the correlation length, $\xi/\xi_0
\sim 2$. 

 We note that the overall scale for $\zeta/s$ is set by the 
two parameters $\eta_0/s$ and $s\xi_0^3$. While $\eta_0/s$ is fairly
well constrained by data, this is not the case for $s\xi_0^3$. 
We have assumed that $s\xi_0^3\simeq 1$, based on the idea that in a 
hadron gas the correlation length cannot be shorter than the 
distance between particles. However, the exact value of $\xi_0$ 
will depend on the composition of the gas, and the precise 
observable used to define the correlation length. For example, 
the recent work of Akamatsu et al.~\cite{Akamatsu:2018vjr} uses 
$\xi_0\simeq 1.2$ fm. Combined with a freezeout value $sT^3\simeq 7$ 
\cite{Cleymans:2005xv} this number corresponds to  $s\xi_0^3\simeq 5$. 

 We have also studied the response function for the bulk pressure
in the critical regime. As expected, relaxation is very slow. 
The bulk pressure approaches its equilibrium value as $1/\sqrt{t}$, 
and the initial rise of the bulk pressure involves a much smaller
power of the correlation length than the scaling of $\zeta$. Note 
that in the present work we have only considered the Fourier transform 
of the equilibrium response. A more complete treatment in an expanding 
system would be based on computing the response from the real time 
correlation functions of the order parameter in an expanding medium 
\cite{Akamatsu:2018vjr,Martinez:2018wia}.

Finally, we may compare our results to simple scaling relations
that have been discussed in the literature. Weinberg proposed
the relation $\zeta \sim (c_s^2-1/3)^2 \eta$, where $c_s$ is
the speed of sound \cite{Weinberg:1972}. This relation is
satisfied in the weak coupling limit of QCD \cite{Arnold:2006fz},
and it holds in simple relaxation time models of kinetic
theory \cite{Dusling:2011fd}. Weinberg's relation predicts
an enhancement of the bulk viscosity near the critical point,
but no critical divergence of the type studied in this work.
Based on an analysis of spectral sum rules, Karsch et al.~suggested
a different scaling relation, $\zeta/s \sim (1/c_s^2-3)$
\cite{Karsch:2007jc}. This formula predicts a mild divergence
of the bulk viscosity, $\zeta/s\sim \xi^{\alpha/\nu}$, much
weaker than our result $\zeta/s\sim \xi^{z-\alpha/\nu}$. 

Implementations of second order non-conformal relativistic fluid
dynamics take into account a bulk viscous relaxation time $\tau_\zeta$.
A typical assumption, motivated by kinetic theory, is that $\tau_\zeta
\sim \zeta/P$ \cite{Dusling:2011fd,Monnai:2016kud}. This relation
captures in an approximate way the critical slowing down of the
response discussed in Sect.~\ref{sec_exp}, but a simple relaxation
time approximation cannot capture the hydrodynamic tail given in
equ.~(\ref{zeta_tail}).

We would like to thank M.~Stephanov and Yi Yin for useful discussions. 
This work was supported in part by the US Department of Energy grant
DE-FG02-03ER41260 and by the BEST (Beam Energy Scan Theory) DOE Topical
Collaboration.

\appendix

\section{Ising equation of state}
\label{sec_ising_eos}

We use a parameterization of the Ising equation of state constructed
by Zinn-Justin \cite{Zinn-Justin:2002ru}. We write
\be
G[\psi,r] = h_0M_0 R^{2-\alpha} g(\theta)
\ee
with
\be
g(\theta)= g(1) + c_1\left(1-\theta^2\right)
+  c_2\left(1-\theta^2\right)^2
+  c_3\left(1-\theta^2\right)^3\, ,
\ee
and $g(1)=0.0424455$ as well as 
\be
c_1=0.321329    \, , \hspace{0.5cm}
c_2=-1.20375    \, , \hspace{0.5cm}
c_3=-0.00126    \, . 
\ee
The magnetic equation of state is $h=h_0R^{\beta\delta}\tilde{h}(\theta)$
with
\be
\tilde{h}(\theta) = \theta+h_1\theta^3+h_2\theta^5
\ee
and
\be
h_1=0.76201\, , \hspace{0.5cm}
h_2=0.00804\, . 
\ee


\end{document}